\documentclass{aastex62}

\def\teff{${\rm T_{eff}}$}

\graphicspath{{./}{figures/}}

\shorttitle{AGB stars in NGC~6752}
\shortauthors{Mucciarelli et al.}

\begin{document}
\title{CONFIRMING THE PRESENCE OF SECOND POPULATION STARS AND THE {\em IRON DISCREPANCY} ALONG THE AGB OF THE GLOBULAR CLUSTER NGC~6752
\footnote{Based on observations collected at the ESO-VLT under the programs 
073.D-0211 and 095.D-0320.}}

\correspondingauthor{Alessio Mucciarelli}
\email{alessio.mucciarelli2@unibo.it}

\author{A.Mucciarelli}
\affil{Dipartimento di Fisica e Astronomia, Universit\`a degli Studi di Bologna, Via Gobetti 93/2, I-40129 Bologna, Italy}
\affil{INAF - Osservatorio di Astrofisica e Scienza dello Spazio di Bologna, Via Gobetti 93/3, I-40129 Bologna, Italy;}

\author{E.Lapenna}
\affil{Dipartimento di Fisica e Astronomia, Universit\`a degli Studi di Bologna, Via Gobetti 93/2, I-40129 Bologna, Italy}
\affil{INAF - Osservatorio di Astrofisica e Scienza dello Spazio di Bologna, Via Gobetti 93/3, I-40129 Bologna, Italy;}

\author{C.Lardo}
\affil{Laboratoire d'astrophysique, Ecole Polytechnique F{\`e}d{\`e}rale de Lausanne (EPFL), 
Observatoire de Sauverny, 1290, Versoix, Switzerland}

\author{P.Bonifacio}
\affil{GEPI, Observatoire de Paris, Universit{\'e} PSL, CNRS, Place Jules
Janssen, 92195 Meudon, France}

\author{F.R.Ferraro}
\affil{Dipartimento di Fisica e Astronomia, Universit\`a degli Studi di Bologna, Via Gobetti 93/2, I-40129 Bologna, Italy}
\affil{INAF - Osservatorio di Astrofisica e Scienza dello Spazio di Bologna, Via Gobetti 93/3, I-40129 Bologna, Italy;}

\author{B.Lanzoni}
\affil{Dipartimento di Fisica e Astronomia, Universit\`a degli Studi di Bologna, Via Gobetti 93/2, I-40129 Bologna, Italy}
\affil{INAF - Osservatorio di Astrofisica e Scienza dello Spazio di Bologna, Via Gobetti 93/3, I-40129 Bologna, Italy;}



\begin{abstract}

Asymptotic giant branch (AGB) stars in the globular cluster NGC6752 have been found to exhibit some chemical peculiarities with respect to the red giant branch (RGB) stars. A discrepancy between [FeI/H] and [FeII/H] (not observed in RGB stars) has been detected adopting spectroscopic temperatures. Moreover, a possible lack of second-population stars along the AGB was claimed.
The use of photometric temperatures based on (V-K) colors was proposed to erase this {\sl iron discrepancy}. 
Also, {\sl ad hoc} scenarios have been proposed to explain the absence of second-population AGB stars.

Here we analyzed a sample of 19 AGB and 14 RGB stars of NGC6752 observed with the spectrographs UVES.
The two temperature scales agree very well for the RGB stars while for the AGB stars 
there is a systematic offset of $\sim$100 K.
We found that even if the photometric temperatures alleviate the {\sl iron discrepancy} with respect to the spectroscopic ones, 
a systematic difference between [FeI/H] and [FeII/H] is still found among the AGB stars. An unexpected result is that the photometric temperatures do not satisfy the excitation equilibrium in the AGB stars.
This suggests that standard 1D-LTE model atmospheres are unable to properly describe the thermal structure of AGB stars, 
at variance with the RGB stars.

The use of photometric temperatures confirms the previous detection of second-population AGB stars in this cluster, with the presence of clear correlations/anticorrelations among the light element abundances. 
This firmly demonstrates that both first and second-population stars evolve along the AGB of NGC6752.

\end{abstract}

\keywords{globular clusters: individual (NGC~6752) --- stars: abundances ---
  stars: AGB and post-AGB --- techniques: spectroscopic}


  \section{Introduction}
Two main results obtained in recent years have revived the interest in the chemical composition 
of asymptotic giant branch (AGB) stars in globular clusters (GCs).
Firstly, is the discovery that iron abundances derived from neutral and single ionized Fe 
lines systematically differ in AGB stars
\citep{ivans01,lapenna14,lapenna15,mucciarelli15a,mucciarelli15b}. 
In particular, Fe~I lines provide lower abundances (by 0.15-0.25 dex) 
with respect to Fe~II lines, only the latter providing abundances consistent with those 
measured in red giant branch (RGB) stars of the same cluster.
This {\em iron discrepancy} has not been observed among the RGB stars, where the two 
sets of Fe lines provide consistent abundances. A qualitative explanation 
of this discrepancy, originally proposed by \citet{ivans01}, is that non-local thermodynamical equilibrium (NLTE) effects, 
which significantly impact neutral lines but only marginally affect single ionized lines, 
are present in the atmospheres 
of AGB stars. However, this interpretation is not fully satisfactory, 
since the NLTE corrections predicted by current theoretical models are similar for AGB and RGB stars
\citep[see e.g.][]{bergemann12,lind12}.

Secondly, it has been speculated that some cluster stars characterized 
by a strong enhancement in N and Na and a depletion in C and O fail to ascend the AGB phase. 
These stars (usually called second population stars, hereafter 2P stars) should have formed from the gas ejected 
by the first stars formed in the cluster (the so-called first population stars, hereafter 1P stars) and 
characterized by light element abundances that well resemble those measured in field stars of similar metallicity. 
It is now well established that all old and massive clusters host a  mixture of 1P and 2P stars,
and the fraction of 2P stars strongly correlates with present-day GC mass \citep[see e.g.][]{gratton12,bastian17}.
However, it has been also observed that the fraction of 2P stars along the AGB is generally smaller than that of the RGB phase in a given cluster. Early 
evidence of this difference is based on the analysis of CN molecular bands in low-resolution spectra \citep{norris81,smith93}, but recent studies based on high-resolution 
spectra confirm this finding 
\citep[see e.g.][]{campbell13,lapenna15,lapenna16,wang17}.
This can be explained by the fact that 2P stars should also be He rich and have a lower mass than 1P stars.
According to standard stellar evolution, stars with masses below 0.55$M_{\odot}$
are expected to skip the AGB phase after the central He-burning phase 
\citep[the so-called {\sl AGB-manqu{\'e}} stars, see e.g.][]{greggio90}.

The existence and the extent of both the iron discrepancy and the dearth of 2P stars along the AGB are still highly debated, with the nearby GC NGC 6752 representing one of the most intriguing cases. 
\citet[][hereafter C13]{campbell13} derived the Na abundance of 20 AGB cluster
stars using GIRAFFE-FLAMES@VLT spectra and concluded 
that they all belong to 1P. Since this
is not expected from standard stellar evolution \citep{cassisi14}, 
a very strong mass-loss in 2P stars during the horizontal branch phase has been invoked by C13 to account for the observed lack of 2P AGB stars. 
However, this assumption is not supported by current models of stellar wind in horizontal 
branch stars \citep{vink02}.

\citet[][hereafter L16]{lapenna16} analyzed the same stars presented in C13 re-observed 
at higher spectral resolution with the spectrograph UVES@VLT. They found that 
(1)~the iron abundances measured from Fe~I lines are lower (by about 0.2 dex) 
than those derived from Fe~II features, confirming the occurrence of the 
{\em iron discrepancy} also in the AGB population of NGC~6752 
(note that C13 did not directly measure iron abundances from their spectra, 
but they assumed a constant Fe abundance from the literature);
and 
(2)~2P stars are present also along the AGB (in contrast with the conclusions 
reached by C13), and the AGB stars show
clear evidence of C-N, Na-O anticorrelations and N-Na, Al-Na correlations.
The analysis by L16 demonstrates that the AGB population of NGC~6752 is composed by a mixture of 1P and 2P stars,  
lacking only the most extreme population (characterized by the highest Na and the 
lowest O abundances) which is observed among the RGB stars of the cluster. This result agrees with the  
expectations from standard stellar evolution \citep[see e.g.][]{cassisi14}
and it has been recently confirmed by \citet{gruyters17} using Str{\"o}mgren photometry.

\citet[][hereafter C17]{campbell17} then questioned the result obtained by L16. They concluded that the {\em iron discrepancy} found by L16 is due to the use of spectroscopic effective temperatures (\teff), while the adoption of   
\teff\ based on the classical infrared flux method
\citep[IRFM, originally proposed by][]{blackwell77}
and the ${\rm (V-K)_0}$   broad-band color,  reconciles
the abundances obtained from Fe~I and Fe~II lines.
Moreover, C17 argued that the light element correlations/anticorrelations  
derived by L16 should be revised in light of the proposed photometric \teff\ scale.

Another still debated case is M4, for which the chemical analyses  
by \citet{maclean16} and \citet{maclean18} suggest a clear lack of 
1P stars along the AGB (at variance with the RGB), while opposite results 
have been obtained by \citet{lardo17}, who found a comparable broadening of AGB 
and RGB sequences using the $C_{UBI}$ index sensitive to the light element abundances, 
and by \citet{marino17}, who detected the Na-O anticorrelation among the AGB stars 
using UVES-FLAMES spectra.

In this paper we present a re-analysis of the spectra of AGB stars in NGC~6752 originally discussed in L16 using 
the \teff\ scales indicated by L16 and C17 to conclusively assess their impact on the measure of Fe~I, Fe~II, C, N, O, Na and Al chemical abundances.

\section{Observational data}
\label{obsdata}
We analyzed two different spectroscopic datasets of stars in NGC~6752:\\ 
{\sl AGB sample}---we used the high-resolution spectra collected with 
the UVES spectrograph \citep{dekker00} at the ESO-VLT
(under the program 095.D-0320, PI:Mucciarelli)
for 20 AGB cluster stars, already analyzed in L16.
The targets are the same previously discussed in C13 and revised in C17. 
Because C17 derived new photometric \teff\ for only 19 out 20 AGB stars of the 
original sample of C13, in the following we restrict the analysis to those stars only.
The observations have been obtained with the Dichroic1 mode employing 
the gratings 390 Blue Arm CD\#2 and 580 Red Arm CD\#3, and adopting 
the 1 arcsec slit that provides a spectral resolution of 40,000. 
More details on the observations and data reduction can be found in L16.\\ 
{\sl RGB sample}---as a reference sample, we analyzed
archival high-resolution spectra for 14 RGB cluster stars 
secured with UVES-FLAMES@VLT \citep{pasquini00} under program 
073.D-0211 (PI:Carretta). The observations have been performed adopting the setup 580 Red Arm CD\#3 with a 
We refer to \citet{carretta09} for more details on the observations.

\section{Chemical analysis}
\label{chems}
The chemical abundances of Fe, Na and Al have been 
determined by using the code {\tt GALA} \citep{m13g} 
through the measure of the equivalent widths of unblended lines. The equivalent 
widths have been measured using the code {\tt DAOSPEC} \citep{stetson08} 
managed through the wrapper {\tt 4DAO} \citep{4dao}. 
The abundances of C, N and O have been obtained through our own code {\tt SALVADOR} 
that performs a \emph{chi-square} minimization between observed and synthetic spectra, 
the latter calculated with the code {\tt SYNTHE} \citep{sbordone04, kurucz05}.
We refer to L16 for details about the analysis procedure and the selection 
of the used transitions. 

We derived chemical abundances using two sets of atmospheric parameters:
\begin{enumerate}


\item the first set of atmospheric parameters is that obtained by using the
{\sl hybrid method} described in \citet{mucciarelli15a} and already adopted 
by L16 for the AGB sample.
In this approach \teff\ are derived spectroscopically through the excitation equilibrium, 
by flattening the slope between the Fe~I line abundances, and the excitation 
potential, $\chi$. Surface gravities (log~g) have been derived 
through the Stefan-Boltzmann relation, adopting the spectroscopic \teff , 
a distance modulus (m-M)$_{V}$=~13.13 mag \citep{harris10}, a color excess E(B-V)=~0.04 mag
\citep{ferraro99} and stellar masses of 0.80 and 0.61  M$_{\odot}$ 
for RGB and AGB targets, respectively.
Microturbulent velocities (v$_{t}$) have been obtained by requiring no trend between iron abundances 
and the reduced equivalent widths. 
This approach is suitable for high-quality, large spectral coverage spectra 
for which robust spectroscopic \teff\ can be derived, avoiding the risk 
of incorrect spectroscopic log~g in case of NLTE or other systematic
discrepancies between Fe~I and Fe~II lines (that could affect the AGB stars).


\item The second set of parameters adopted here has been obtained with the method 
used by C17  who photometrically derived  \teff\ and log~g. In particular, C17 derived  
\teff\ using the IRFM as implemented by \citet{casagrande10}
and adopting the broad-band ${\rm (V-K)_0}$ color. 
We reanalyzed the AGB target stars adopting the values of \teff\ and log~g calculated 
by C17, while v$_{t}$ have been derived spectroscopically to take advantage 
of the large number of Fe lines available in the UVES spectra 
\citep[at variance with C17, who derived this parameter adopting the log~g-v$_{t}$ relation by][]{gratton96} . \\
For the RGB stars, we derived \teff\ using the ${\rm (V-K)_0}$-\teff\
relation provided by \citet{casagrande10}, while log~g have been 
obtained from the Stefan-Boltzmann relation. As done by C17, we adopted
the optical photometry by \citet{momany02} and the near-infrared 
photometry from the 2MASS database \citep{skrutskie}.
\end{enumerate}

\begin{figure}[ht]
\centering
\includegraphics[clip=true,scale=0.75,angle=0]{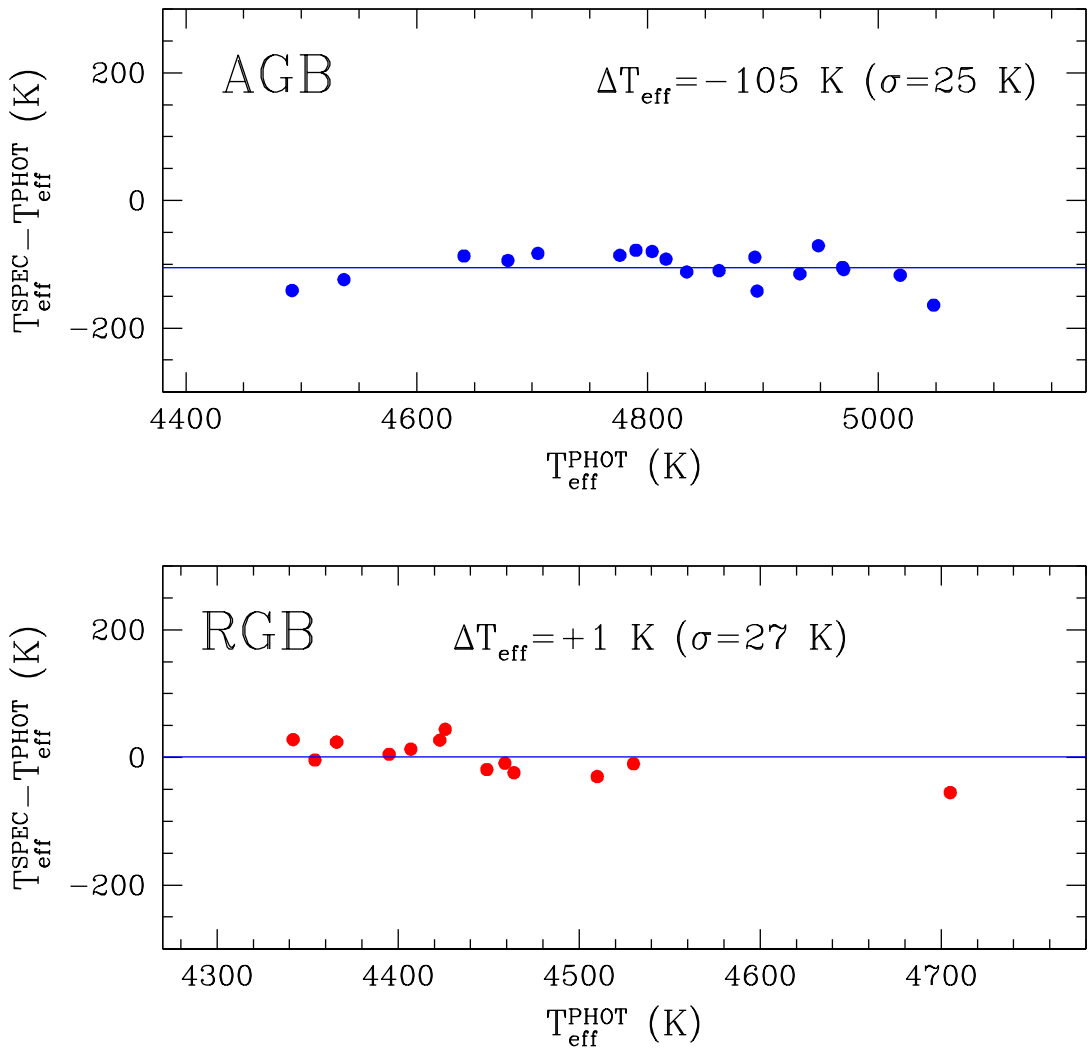}
\caption{Difference between spectroscopic and photometric \teff\ as a function 
of the photometric \teff\ for AGB and RGB samples (upper and lower panel, respectively).
The blue solid lines mark the average difference values, which are also labeled in each panel.}
\label{diff}
\end{figure}

We compare the two sets of adopted \teff\ for AGB and RGB stars, separately.
Fig.~\ref{diff} shows the difference between spectroscopic and photometric 
\teff\ as a function of the photometric \teff\ for the two samples. 
For the AGB stars the average difference is $\Delta$\teff\ = --105 K ($\sigma$=~25 K), 
indicating a systematic offset between the two scales, while 
for the RGB stars the two scales agree very well, with an average difference of only +1 K ($\sigma$=~27 K). 

\section{Results}

\label{res1}

In this section we quantitatively investigate 
the dependence of the derived abundances on the adopted atmospheric parameters discussed in
Section~\ref{chems}.

\begin{figure}[ht]
\centering
\includegraphics[clip=true,scale=0.75,angle=0]{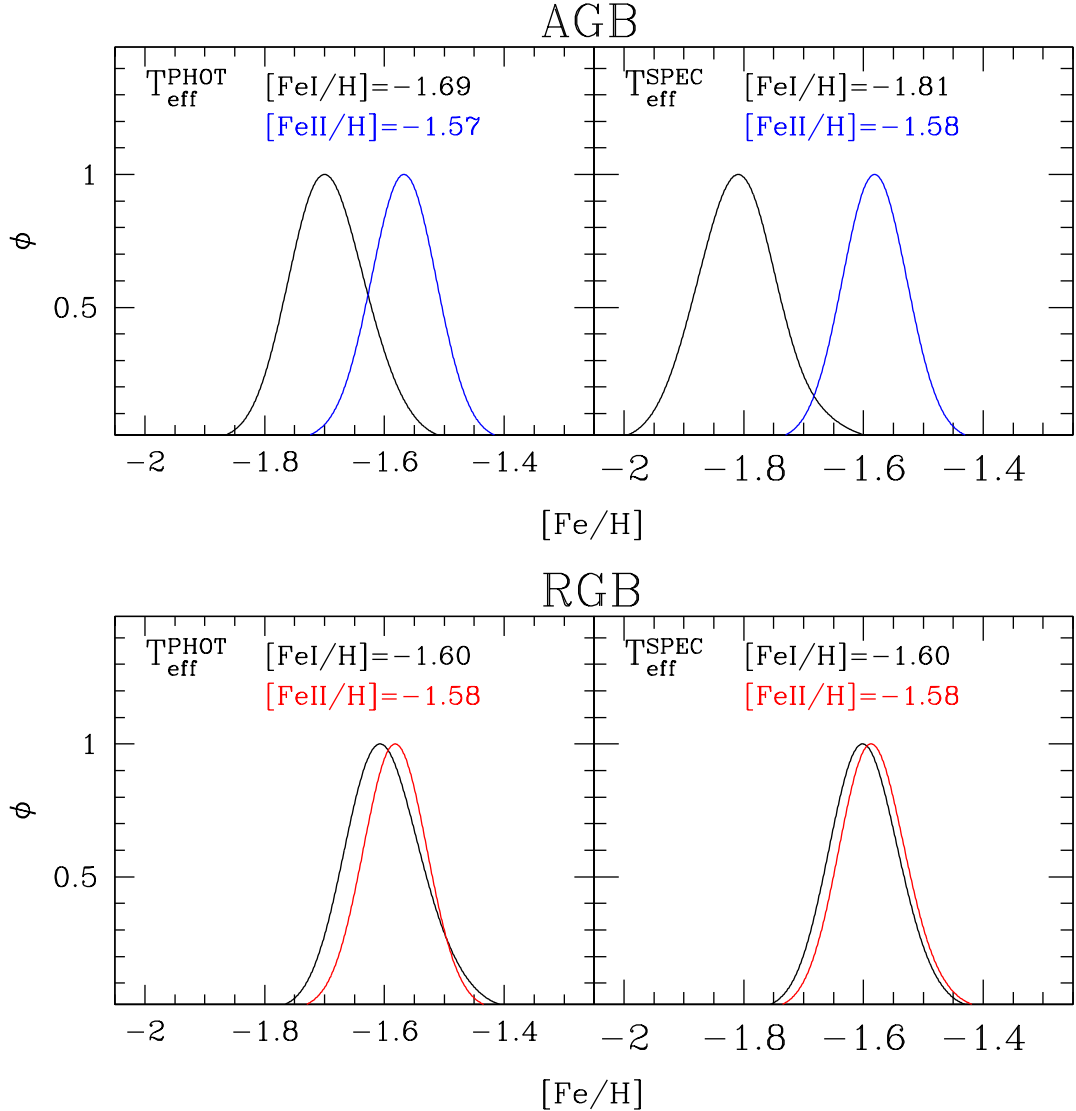}
\caption{[Fe/H] distributions for the AGB and RGB samples 
(upper and lower panels, respectively) as derived from Fe~I 
(black histograms) and Fe~II lines (blue and red histograms). 
The left panels show the [Fe/H] distributions obtained 
with the photometric \teff\ scale used by C17, while the right 
panels display those obtained with the spectroscopic \teff\ .}
\label{iron}
\end{figure}

\begin{figure}[h]
\centering
\includegraphics[clip=true,scale=0.75,angle=0]{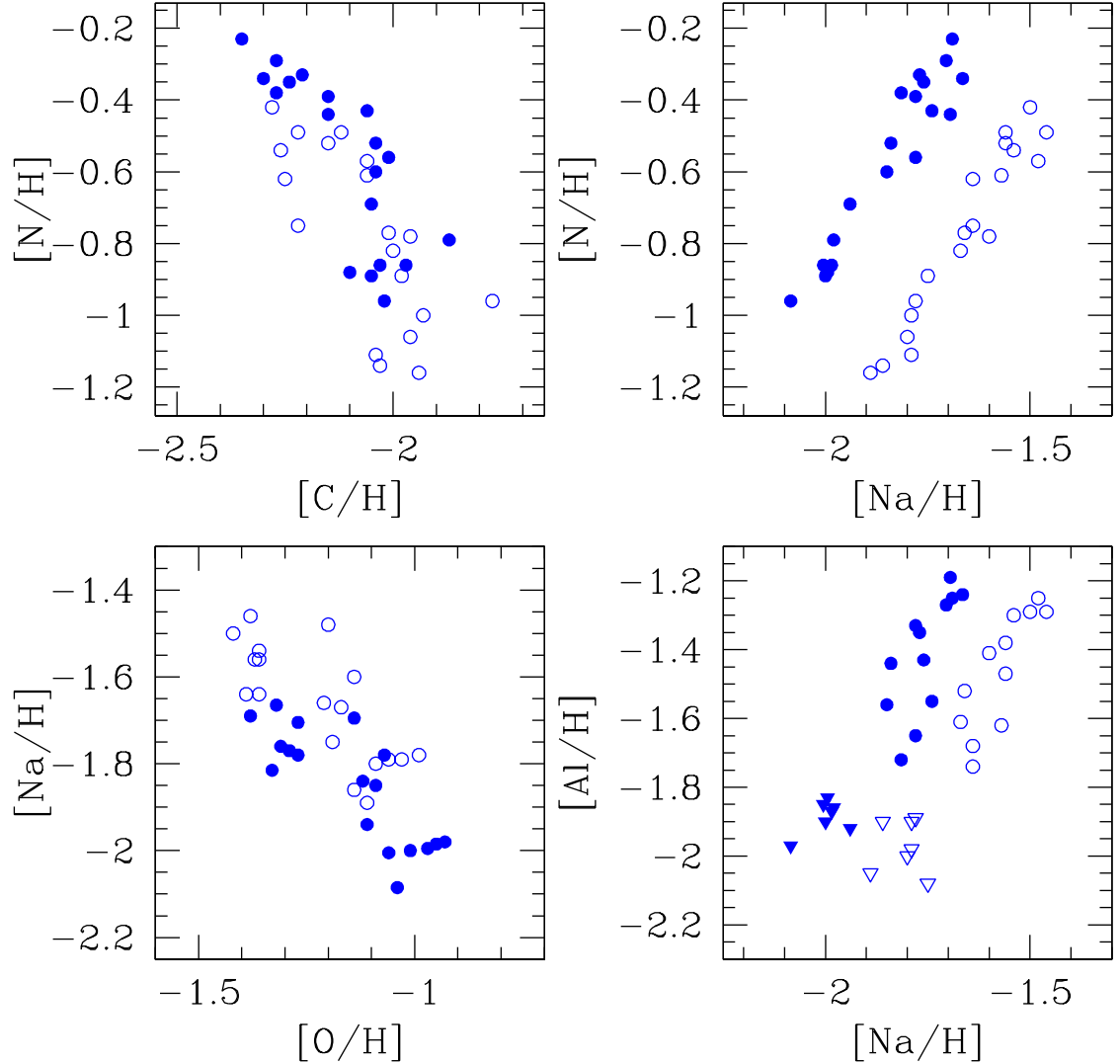}
\caption{Light element abundance ratios (normalized to hydrogen) in the AGB stars of NGC~6752 
calculated with the photometric \teff\ by C17 (blue filled circles) and with the 
spectroscopic \teff\ by L16 (blue empty circles). Reversed triangles indicate 
upper limits for [Al/H].}
\label{light}
\end{figure}

\begin{figure}[h]
\centering
\includegraphics[clip=true,scale=0.75,angle=0]{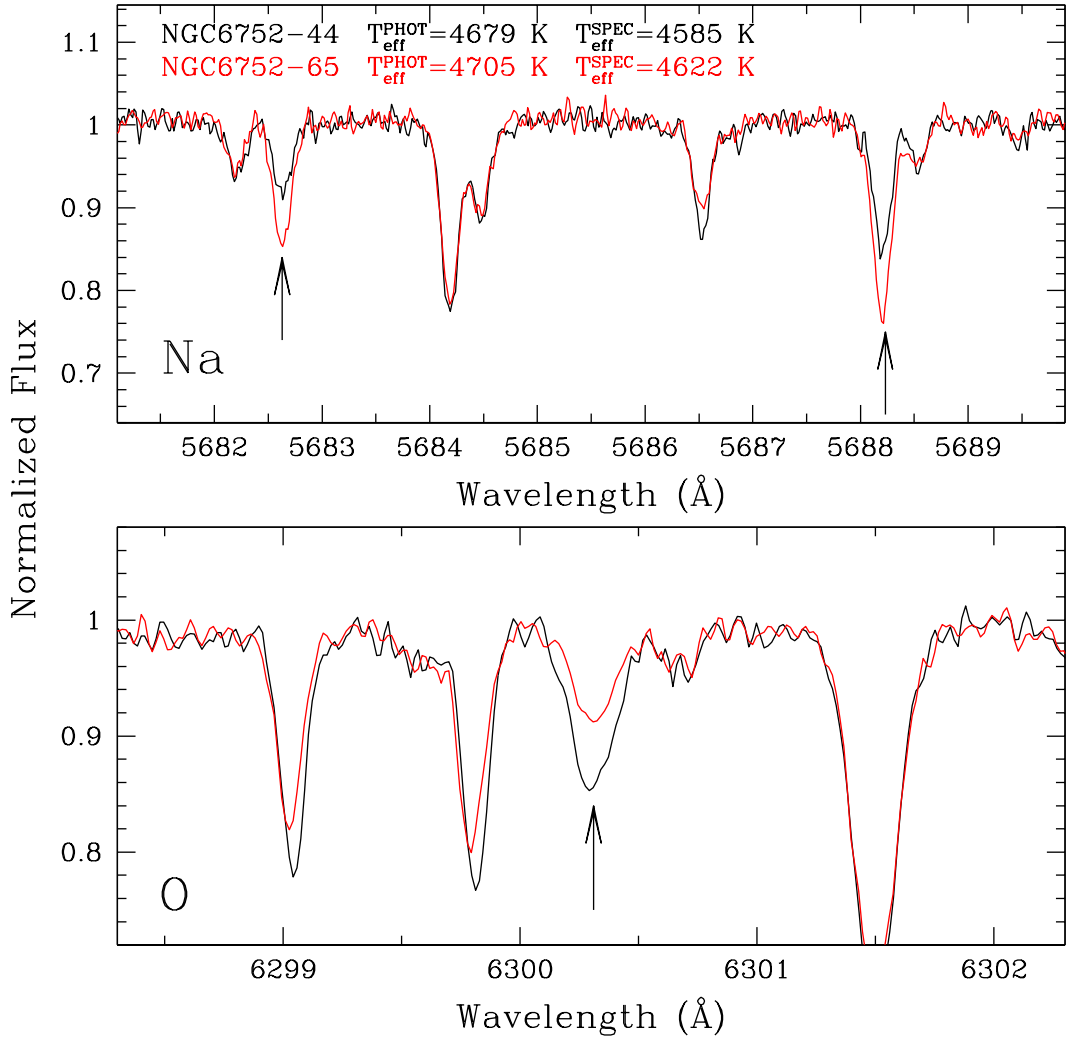}
\caption{Comparison between the spectra of the AGB stars \#44 and \#65 (black and red line, respectively), 
in the spectral regions around the Na~I doublet (upper panel) and the forbidden O~I line (lower panel). 
Photometric and spectroscopic \teff\ for the two targets are labelled in the upper panel.}
\label{compaspec}
\end{figure}

\begin{figure}[h]
\centering
\includegraphics[clip=true,scale=0.75,angle=0]{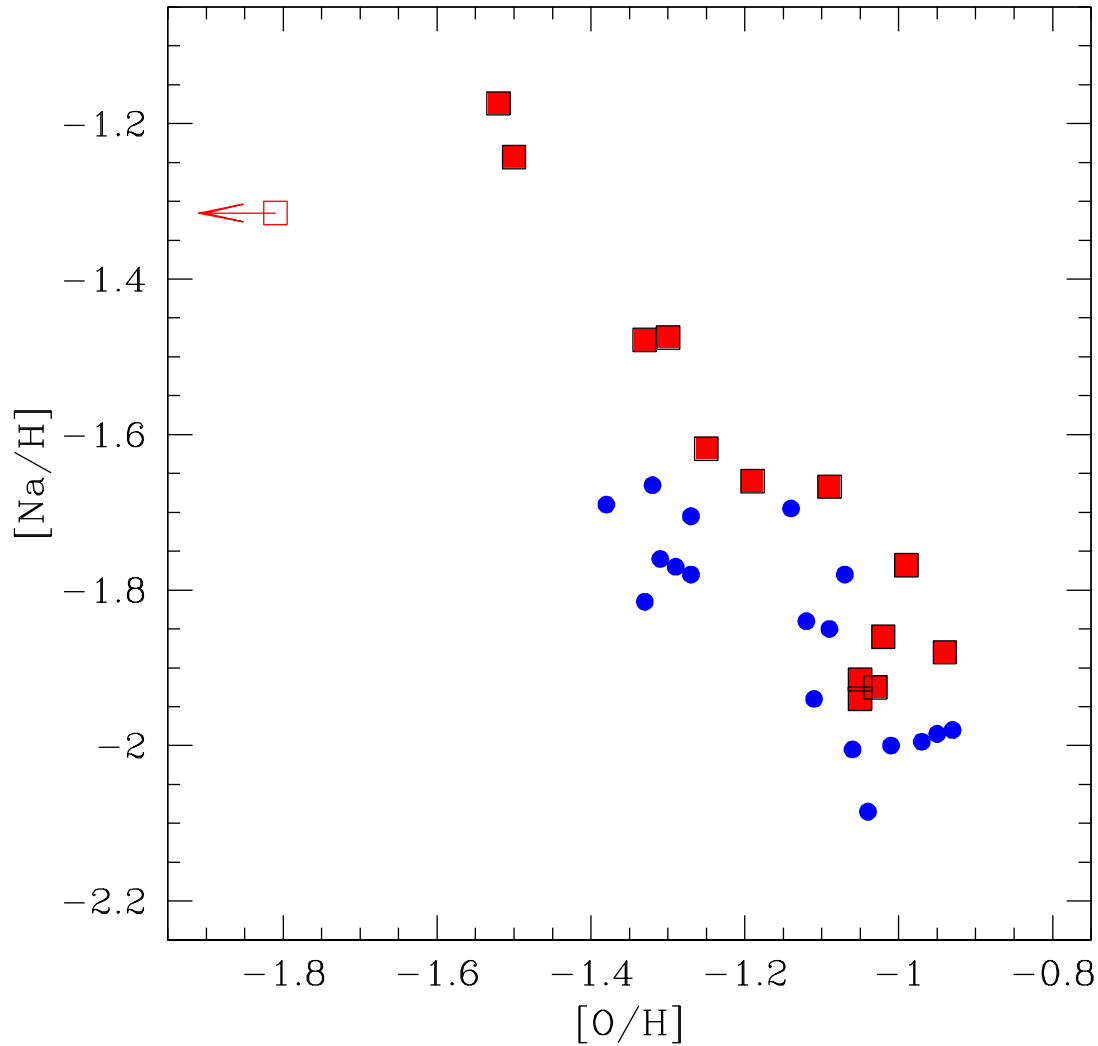}
\caption{Behavior of [Na/H] as a function of [O/H] for the AGB and the RGB stars (blue circles and red squares, respectively) 
measured adopting the photometric \teff . The arrow indicates an upper limit for [O/H].}
\label{naox}
\end{figure}

\begin{enumerate}
\item
{\em Iron discrepancy}---
Table 1 reports the average Fe abundances from neutral and single ionized lines 
for AGB and RGB stars obtained by using the two \teff\ scales.
Fig.~\ref{iron} shows the Fe~I and Fe~II metallicity distributions
for the two stellar samples, obtained by adopting the 
photometric (left panels) and the spectroscopic (right panels) \teff\ scales. 
The metallicity distributions are shown as a generalized histograms, 
a representation that removes the effects due to the choice of the starting 
point and of the bin size by taking  
the uncertainties in each individual [Fe/H] value into account \citep{laird88}.
\\
For the RGB stars photometric and spectroscopic \teff\ agree very well (see Fig.~\ref{diff}), 
leading to very similar iron abundances both from Fe~I and Fe~II lines. 
In the case of the AGB stars, a significant difference remains between 
[Fe~I/H] and [Fe~II/H] regardless of the adopted \teff . 
In particular, the difference between [Fe~I/H] and [Fe~II/H] is of --0.23 dex 
if spectroscopic \teff\ are assumed,  and --0.12 dex with the photometric \teff .
The higher values of \teff\ derived from C17 lead to an increase 
of [Fe~I/H] (a change of $\pm$100 K leads to a variation of $\pm$0.12 dex in [Fe~I/H]).
We also stress that Fe~II lines in AGB stars provide the same abundance 
found for RGB stars from both neutral and single ionized lines, regardless of the 
adopted \teff , 
{\sl confirming that Fe~II lines are the most reliable indicator 
of metallicity for these stars}. \\
As a further check, we reanalyzed also the archival GIRAFFE spectra used by C13 and C17, 
adopting the atmospheric parameters derived by C17 and a suitable linelist including 
Fe lines predicted to be unblended according to the cluster metallicity, the stellar parameters 
and the spectral resolution of GIRAFFE. For AGB stars we derived [Fe~I/H]=--1.68$\pm$0.01 dex ($\sigma$=~0.05 dex) 
and [Fe~II/H]=---1.56$\pm$0.01 dex ($\sigma$=~0.03 dex). For the RGB stars observed by 
C13 and C17 (a different sample 
with respect to the reference one analyzed here and described in Section~\ref{obsdata}) we derived 
[Fe~I/H]=--1.58$\pm$0.01 dex ($\sigma$=~0.04 dex) and [Fe~II/H]=--1.55$\pm$0.01 dex ($\sigma$=~0.05 dex), 
in perfect agreement with the results 
obtained from the two UVES samples.

\item
{\sl Light element abundances}---
Fig.~\ref{light} shows the C-N and O-Na anticorrelations and 
the N-Na and Na-Al correlations obtained for the AGB stars of NGC~6752 using 
the spectroscopic \teff\ scale by L16 (blue empty circles) and the photometric one of C17 (blue filled circles). 
The two sets of abundance ratios (normalized to hydrogen) exhibit 
the same patterns. 
The difference of about 100 K between the two \teff\ scales leads to a systematic 
offset in the abundance ratios without changing the chemical patterns already detected by L16 with the spectroscopic \teff\ .

L16 show in their Figure 1 the comparison between the spectra of two AGB cluster stars 
(namely \#44 and \#65) 
with similar parameters but different depths concerning Na, O and Al atomic lines and 
CN and NH molecular bands, and similar depths for the other metallic lines. 
In Fig.~\ref{compaspec} we show the spectral regions around the Na~I doublet at 5682-88 \AA\ (upper panel) 
and the forbidden O~I line at 6300.3 \AA\ (lower panel) for these two stars.
C17 provide for these two stars very similar photometric \teff\  
(with a difference of 26 K only). Hence, the use of their \teff\ scale cannot 
explain the different depths of these molecular and atomic lines, which can be attributed only to 
an intrinsic chemical abundance difference. 

Fig.~\ref{naox} shows the trend between [Na/H] and [O/H] for the AGB (blue circles) and 
the RGB (red squares) stars when the photometric \teff\ is adopted. 
For [O/H]$>$--1.4 dex, an offset 
is found between the Na abundances of the two groups of stars, with [Na/H] in the AGB stars being lower by --0.1 dex 
than that measured in the RGB stars.
The origin of this small offset is unclear but it cannot be attributed to systematics in the analysis, 
because we adopted the same linelist, solar values, \teff\ scales and NLTE corrections 
\citep[the latter from][]{lind11} for the two stellar samples. 
Despite this offset, the two samples exhibit a clear Na-O anticorrelation, showing a different extent. In agreement with the results of L16, objects with the highest Na 
and the lowest O abundances observed among the RGB stars are missing along the AGB 
\citep[see also][]{carretta09,yong03}.

Hence, also the light element abundances obtained by using the photometric \teff\  support the conclusion by L16: 
the AGB stars in NGC~6752 include a mixture of 1P and 2P stars. 
Indeed, the existence of clear chemical patterns among the light element abundances of AGB stars
is not compatible with the presence of 1P stars only. 
This result agrees with theoretical predictions for NGC~6752 \citep{cassisi14} 
and with the evidence based on Str{\"o}mgren photometry provided by \citet{gruyters17} 
that demonstrates the presence of three sub-populations in the RGB of the cluster but 
only two sub-populations in the AGB. In particular, the position of the 1P and 2P stars 
as identified by L16 according to their Na and O abundances well correlates with the 
two photometric branches observed along the AGB.

\end{enumerate}

\begin{table*}
  \begin{center}
  \caption{Average [Fe~I/H] and [Fe~II/H] abundance ratios for the AGB and RGB samples, adopting photometric and spectroscopic \teff\ .}
  \label{abu}
  \begin{tabular}{lcccc}
  \hline
  
  \teff\ &  [Fe~I/H]$_{\rm RGB}$  &  [Fe~II/H]$_{\rm RGB}$  & [Fe~I/H]$_{\rm AGB}$  & [Fe~II/H]$_{\rm AGB}$ \\
  
  \hline   
  
     {\rm PHOT}   &     --1.60$\pm$0.01 ($\sigma$=0.04)   &      --1.58$\pm$0.01 ($\sigma$=0.02)  &      --1.69$\pm$0.01 ($\sigma$=0.04)   &   --1.57$\pm$0.01 ($\sigma$=0.02)   \\  
     {\rm SPEC}     &     --1.60$\pm$0.01 ($\sigma$=0.03)   &      --1.58$\pm$0.01 ($\sigma$=0.03)  &      --1.81$\pm$0.01 ($\sigma$=0.05)    &    --1.58$\pm$0.01 ($\sigma$=0.02)  \\

\hline 
\end{tabular} 
\end{center}
\end{table*}

\section{AGB vs RGB: some missing physics?}

The comparison between the spectroscopic and ${\rm (V-K)_0}$-based \teff\ 
in RGB and AGB stars of NGC~6752 provides an unexpected result.
As discussed in Section~\ref{res1}, photometric and spectroscopic \teff\ agree 
very well in RGB stars but they are different by $\sim$100 K in AGB stars.
The spectra of the two stellar samples are very similar in terms of spectral coverage, 
spectral resolution and signal-to-noise ratio. Also, the analysis of AGB and RGB stars 
is based on the same linelist and the same model atmospheres. Hence, this different 
behavior cannot be attributed to some systematics in the analysis.

\begin{figure}[h]
\centering
\includegraphics[clip=true,scale=0.75,angle=0]{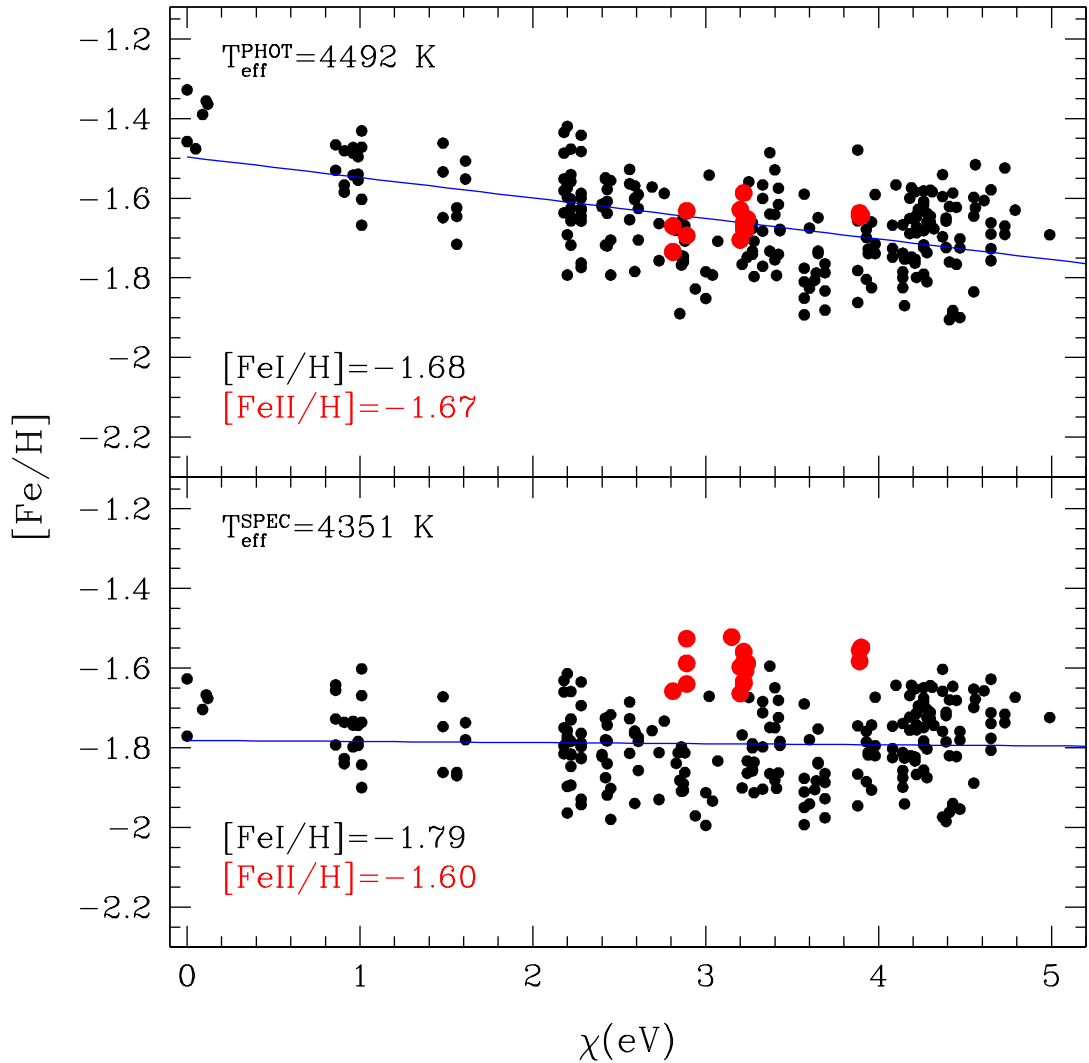}
\caption{Behavior of the iron abundances as a function 
of the excitation potential, $\chi$, for the AGB star \#25, in the case of the photometric 
\teff\ adopted by C17 (upper panel) and of the spectroscopic \teff\  
derived by L16 (lower panel). 
The black circles are for the abundances derived from neutral Fe lines, the red circles are for the single ionized Fe lines. The blue 
lines are the best linear fits obtained to the Fe~I abundances.}
\label{eptrend}
\end{figure}

Fig.~\ref{eptrend} shows the behavior of the iron abundances as a function 
of $\chi$ for the AGB star NGC~6752-25, for which a difference of about 150 K is found between 
photometric and spectroscopic \teff . 
While a good agreement between [Fe~I/H] and [Fe~II/H] is derived using the photometric \teff , 
this temperature scale introduces a significant slope between [Fe~I/H] and $\chi$. 
This trend is canceled out by adopting a cooler \teff\, but this also increases the  
difference between the abundances from Fe~I and Fe~II lines, thus increasing the {\em iron discrepancy}. 
Note that the spectra used by C17 have a small number of Fe~I lines 
(and without lines with $\chi<$2 eV) and they are not suitable to highlight that photometric \teff\ do not satisfy the excitation equilibrium.

The only way to satisfy the excitation equilibrium adopting the photometric 
\teff\ is to increase v$_{t}$ by 0.3-0.5 km/s (changes in log~g do not impact on the 
slope between [Fe~I/H] and $\chi$). However, this choice has two disadvantages: 
it introduces 
a significant, negative slope between [Fe~I/H] and the reduced equivalent width
(i.e. the stronger lines provide systematically lower abundances), pointing out 
that these v$_{t}$ are wrong,
and the derived average iron abundances (both from Fe~I and Fe~II lines) are $\sim$0.1 dex lower 
than those obtained for RGB stars.
In other words, there is no way for the AGB stars to satisfy all the spectroscopic constraints adopting photometric \teff\ (at variance with the RGB stars).

As an additional check, we analyzed the AGB stars with the {\sl hybrid method}
(see Section~\ref{chems}) excluding 
Fe~I lines with $\chi<$2 eV that are the most sensitive to \teff\ and to 3D and NLTE effects 
\citep[see e.g.][]{collet07,mashonkina13,amarsi16}.
Also with this selection, a significant slope between [Fe~I/H] and $\chi$ is found, 
and it can be flattened only decreasing 
\teff\ with respect to the photometric values. The average difference between the spectroscopic 
\teff\ derived including all the lines and and those obtained by using only the high-$\chi$ ones is of --40 K ($\sigma$=~37 K). 
In addition, a discrepancy between the two iron abundances in AGB stars remains, 
with [Fe~I/H]=--1.79$\pm$0.01 dex ($\sigma$=~0.07 dex) and [Fe~II/H]=--1.62$\pm$0.01 dex ($\sigma$=~0.03 dex), 
while no significant difference is found for the RGB stars when the low-$\chi$ lines are excluded
([Fe~I/H]=--1.61$\pm$0.01 dex, $\sigma$=~0.02 dex, [Fe~II/H]=~--1.58$\pm$0.01 dex, $\sigma$=~0.03 dex).

The difference between the two \teff\ scales has been already discussed 
by C17, who suggest that it is due the tendency of the spectroscopic \teff\ to remain close 
to the initial guess value (in other words, if the prior for \teff\ is incorrect, also the derived 
spectroscopic \teff\ will be incorrect).
However, the GIRAFFE-FLAMES spectra analyzed by C13 and C17 have been acquired with two gratings (HR11 and HR13) that do not 
guarantee a robust determination of the spectroscopic \teff ,
because of the limited spectral coverage and the low number of available lines.
Using the same archival data analyzed in C13 and C17, we noted that all the unblended
and usable Fe~I lines available in the two gratings have excitation potentials higher than $\sim$2 eV.
The lack of low-$\chi$ Fe~I lines (that are the most sensitive to \teff\ ), combined 
with a relatively small number of Fe~I lines (less than 40, compared with more than 200 lines 
available in the UVES spectra)
makes highly uncertain the spectroscopic determination of \teff\ . 
This explains why C17 concluded that
{\sl "the spectroscopically determined temperatures tend to lie 
close to the initial estimates"}. 
We verified what happens in the case of UVES spectra by adopting different starting values of \teff\ .  
For each star, we find that the resulting spectroscopic \teff\ converge to very similar values 
(with changes of less than $\pm$20 K) regardless of the starting value.

The fact that the two \teff\ scales agree in RGB stars but not in AGB stars is 
unexpected and not easy to explain. The two groups of stars have the same 
metallicity  since they belong to the same cluster and the difference in atmospheric 
parameters is not so large to justify this finding.
This seems to suggest that the standard treatment of model atmospheres and line transfer
is unable to properly reproduce the thermal structure of AGB stars (at variance with the RGB 
stars where no significant problem is found). It is hard to say which \teff\ is correct 
for the AGB stars.
If we assume that photometric \teff\ are correct, we need to explain why the excitation equilibrium 
is not satisfied in AGB stars. On the other hand, if we rely on the spectroscopic ones, 
we need to explain why \teff\ based on the ${\rm (V-K)_0}$  colors (a standard and 
well-reliable temperature indicator) provide discrepant results between AGB and RGB stars.
The hypothesis that 3D and/or NLTE effects are larger in AGB stars with respect to 
RGB stars 
cannot be totally ruled out and, indeed, it might also account for
the small offset in [Na/H] that we found between AGB and RGB stars 
(see Fig.~\ref{naox}).

\section{Summary and conclusions}

The comparison between the chemical abundances in the AGB and RGB stars of NGC~6752 
obtained adopting the photometric \teff\ by C17 and the spectroscopic \teff\ by L16 
provides the following results:
\begin{itemize}
\item
the two \teff\ scales agree very well for the RGB stars while for the AGB stars 
a systematic offset of $\sim$100 K does exist. In particular, the photometric \teff\ 
do not satisfy the excitation equilibrium for AGB stars (at variance with the RGB stars). 
In order to flatten the slope between [Fe~I/H] and $\chi$, the photometric \teff\ should be lowered;
\item
the adoption of the photometric \teff\ alleviates the {\em iron discrepancy} in AGB stars but it does not totally erase the difference between [Fe~I/H] and [Fe~II/H] 
(which decreases from 0.23 dex to 0.12 dex), while for RGB stars the 
{\em iron discrepancy} is not found;
\item
the use of photometric \teff\ does not alter the correlations and anticorrelations 
found by L16 among the light elements (C, N, O, Na, Al) confirming that 
both 1P and 2P stars are observed along the AGB of NGC 6752.
This confirms the results of L16, while it is at odds with the conclusions of C13 
and C17;
\item
the use of high-resolution spectra (as GIRAFFE-FLAMES) with a relatively small spectral coverage (hence with a low number of Fe lines)
should be avoided in the study of AGB stars because it does not allow to properly 
check the occurrence of possible correlations between [Fe~I/H] and $\chi$;
\item
the failure of photometric \teff\ to satisfy the excitation equilibrium 
in AGB stars (but not in RGB stars) seems to suggest 
that current model atmospheres are not adequate to properly reproduce the complex thermal structure of these stars.
In this case, neither the photometric nor the spectroscopic \teff\ can be considered 
reliable. In light of these results, Fe~II lines are the most robust metallicity indicators 
for the AGB stars.
\end{itemize}

The {\em iron discrepancy} in AGB stars remains an open problem that calls for 
new and deep investigations, using high-resolution, high-quality 
spectra for the chemical analysis and an effort to better understand the structure of the photospheres of these stars.

\acknowledgements
We thank the anonymous referee for his/her useful comments.
We thank Y. Momany to share with us the photometric catalog of NGC~6752.
CL acknowledges financial support from the Swiss National Science Foundation (Ambizione grant PZ00P2\_168065).

\software{GALA \citep{m13g}, DAOSPEC \citep{stetson08}, 4DAO\citep{4dao}, 
SYNTHE \citep{sbordone04,kurucz05}}




\end{document}